\documentclass[aps,showpacs,nofootinbib]{revtex4-2}
\usepackage{graphicx}
\usepackage{bm}
\usepackage{graphicx}
\usepackage{amsmath}
\usepackage{mathrsfs}
\usepackage{ulem}
\usepackage{color}
\usepackage[colorlinks, linkcolor=red,anchorcolor=green,citecolor=blue]{hyperref}

\newcommand{\nn}{\nonumber}
\newcommand\diag{\operatorname{diag}}

\graphicspath{{./figs/}}

\begin{document}  
	
\title{Normal mode analysis within a mutilated relaxation time approximation}

\author{Jin Hu}
\email{hu-j23@fzu.edu.cn}
\affiliation{Department of Physics, Fuzhou University, Fujian 350116, China}

\begin{abstract}
In this paper, we present a detailed analysis of normal modes based on the Boltzmann equation within  the mutilated relaxation time approximation (RTA). Using this linearized effective kinetic description, our analysis encompasses a complete order calculation in wavenumber $k$, extending the conventional hydrodynamic mode analysis to intermediate and short-wavelength regions. Furthermore, our linear mode analysis can provide a natural classification of kinetic modes into collective modes and non-collective single-particle excitations.
In the case of an energy-independent relaxation time, the  behavior of hydrodynamic onset transitions is recovered \cite{Romatschke:2015gic}. 
However, for the  case with an energy-dependent relaxation time, 
the distinct classification becomes less clear, as the location of hydrodynamic modes is not well separated from  non-hydrodynamic modes.
\end{abstract}




\maketitle	
\section{Introduction}
As a universal effective theory for large distance and time scales, relativistic hydrodynamics has been extensively applied in high energy physics \cite{Romatschke:2017ejr} and cosmology \cite{McDonough:2020tqq}. Recently, relativistic hydrodynamics has found successful application in characterizing the evolution of the fireball generated during relativistic heavy-ion collisions. It has been instrumental in deducing the properties of Quantum Chromodynamics (QCD) matter, quark-gluon plasma (QGP),  through the analysis of experimental data generated at BNL-RHIC and CERN-LHC~\cite{Hama:2004rr,Huovinen:2006jp,Jeon:2015dfa,Bernhard:2019bmu,Auvinen:2020mpc,Nijs:2020roc,JETSCAPE:2020mzn,Parkkila:2021tqq}. Surprisingly, the hydrodynamic model has shown unexpected success in describing the dynamic evolution of the collision systems, such as nucleus-nucleon and proton-proton collisions \cite{CMS:2016fnw,ATLAS:2017hap}, which raises a thought-provoking question of when and how the hydrodynamic behavior emerges in a non-equilibrium system. 

The heavy-ion collision community has extensively employed hydrodynamics as a quantitative tool for interpreting experimental results. Nonetheless, there remains a necessity for a comprehensive understanding of the range of its application. In the phenomenological modeling of relativistic heavy-ion collisions, the system is required to reach rapid equilibration to accurately simulate the evolution of the created QGP and to match the experimental data. Nevertheless, it remains an open question why thermal equilibrium is achieved so quickly in relativistic heavy-ion collisions. Fundamentally, understanding the establishment of thermalization in a physical system that is initially far from equilibrium is a crucial question in various fields. Recently, the attractor structure, familiar in the realm of complex physics, has been recognized within the framework of relativistic hydrodynamics. This discovery provides a reasonable qualitative explanation for the effectiveness of hydrodynamic models even on short time scales. Consequently, considerable effort has been directed towards deciphering the relationship between attractor behavior and early-time dynamics \cite{Heller:2015dha,Strickland:2018ayk,Denicol:2020eij,Behtash:2017wqg,Romatschke:2017vte,Chattopadhyay:2019jqj,Kurkela:2019set}.


As one important step towards understanding how relativistic nonequilibrium systems  thermalize, the research on the  retarded correlators has recently drawn  extensive attention. The late-time behavior of an off-equilibrium system is typically governed by the non-analytical structures within these correlators. For instance,  poles signify the hydrodynamic collective excitations, while the existence of cuts or non-hydrodynamic modes is related to the application range of hydrodynamics. It has been shown that the correlators contain only poles or quasinormal modes at infinite t’Hooft coupling in large $N$ thermal  $\mathcal{N} =4$ Super Yang-Mills theory \cite{Hartnoll:2005ju,Kovtun:2005ev,Grozdanov:2016vgg}, which  motivates  the similar exploration on the side of a weakly-coupled kinetic theory.   The related research is conducted by Paul Romatschke to study the physical behavior of  thermal correlators using effective kinetic theory  \cite{Romatschke:2015gic}, see also \cite{Bajec:2024jez}, revealing the dominant role of hydrodynamic poles as long-lived modes.  However, there is a debate on the dominant non-analytical structures raised by \cite{Kurkela:2017xis}, where the dominant non-analytical structures are found to be the branch-cuts rather than poles. Subsequent studies, including analytical estimates \cite{strain2010,Gavassino:2024rck} and numerical calculations \cite{Moore:2018mma,Ochsenfeld:2023wxz} support the dominant role of branch-cuts. The seemingly contradictory conclusions  reflect that the retarded correlators possess complex non-analytical structures depending on the details of interactions \cite{Hu:2024tnn}.  Both perspectives are valid, provided the discussion is confined to their respective applicable scenarios.

In this work, we focus on the weakly-coupled system, which serve as an excellent starting point for examining both far-from-equilibrium and near-equilibrium dynamics and for clarifying their intrinsic links within a kinetic framework. Specifically, we utilize the linearized Boltzmann equation within the mutilated relaxation time approximation (we shall come to  the meaning of the term “mutilated” later) to seek its normal mode solutions. Notably, we provide an alternative way to unravel the non-analytical structures within the retarded correlators. As previously mentioned, the validity of a relaxation time approximation relies on the interaction details, and determines the non-analytical structures contained in the retarded correlators in turn. To make progress in this direction, we concentrate on the scenarios where the relaxation time approximation is applicable. In the main text, we give a  detailed normal mode analysis,  which encompasses a complete order calculation in wavenumber $k$, and extends the  conventional hydrodynamical analysis to intermediate and short-wavelength regions.  Given that hydrodynamization is typically conceptualized as the emergence of hydrodynamic modes concomitant with the decay of non-hydrodynamic modes, our research concentrates on these phenomena to foster an understanding of the underlying physics. The  paper is organized as follows.  In Sec.~\ref{lke1},
we give a concise review of  the linearized kinetic equation. In Sec.~\ref{relax}, a transport model within the mutilated relaxation time is constructed, and the advantages of this model over traditional RTA are discussed in detail.
Sec.~\ref{analysis} is devoted to our main analysis. Here, we employ an energy-independent relaxation time and conduct a detailed normal mode analysis. The onset transition behavior  documented in \cite{Romatschke:2015gic,Bajec:2024jez} is recovered. When the same analysis is applied to the case with an energy-dependent relaxation time in Sec.~\ref{dep}, the  inseparability of hydrodynamic modes and non-hydrodynamic modes is revealed. 
Their interplay introduces complexity into the late-time dynamic evolution.  Summary and outlook are given in Sec.~\ref{su}. We have also included some related materials in the appendices. Natural units $k_B=c=\hbar=1$ are used. The metric tensor  is given by $g^{\mu\nu}=\diag(1,-1,-1,-1)$ , while $\Delta^{\mu\nu} \equiv g^{\mu\nu}-u^\mu u^\nu$ is the projection tensor orthogonal to the four-vector fluid velocity $u^\mu$. The abbreviation dP stands for $\int dP\equiv \frac{2}{(2\pi)^3}\int d^4p\, \theta(p^0)\delta(p^2 - m^2)$. In addition, we employ the symmetric shorthand notation $X^{( \mu\nu ) } \equiv(X^{ \mu\nu } + X^{ \nu \mu})/2$.

 \section{Linearized kinetic equation}
 \label{lke1}
The non-equilibrium evolution of  a relativistic system is governed by the relativistic Boltzmann equation,
\begin{align}
\label{Boltzmann}
&(p\cdot\partial +F\cdot\partial_p)f(x,p)=C[f],
\end{align}
where one-particle distribution function $f$, a function of phase space point $(x,p)$, plays a pivotal role in relativistic kinetic theory, from which other physical observables or quantities can be obtained via phase space integration. The symbols $\partial$ and $\partial_p$  denote the partial derivative with respect to the coordinate $x$ and particle momentum $p$, respectively. $F$ is the external force exerted on the particles, which will be set to zero from now on. Here, the collision term $C[f]$ represents the effect of local collisions between particles on the distribution function $f$,
 \begin{align}
 \label{Ckl}
 C[f]
 \equiv\;& 
 \int  {\rm dP^\prime}  {\rm dP_1} {\rm dP_2} \big(f(p_1)f(p_2)\tilde{f}(p)\tilde{f}(p^\prime)-f(p)f(p^\prime)\tilde{f}(p_1)\tilde{f}(p_2)\big) 
 W_{p,p^\prime\to p_1,p_2},
 \end{align}
 where we have invoked the  detailed balance property $W_{p,p^\prime\to p_1,p_2}=W_{p_1,p_2\to p,p^\prime}$ for the transition rate, and $\tilde{f}=1\pm f$ with $-$ for fermions and $+$ for bosons is implied. In the limit of classical statistics, $\tilde{f}$ is trivially $1$, and we will only focus on this case. Only two-body collisions are considered here. Note that \eqref{Boltzmann} is  a one-component version of the Boltzmann equation;  for a detailed discussion of a general multi-component system, see \cite{DeGroot:1980dk, Hu:2022vph}.
 
 Due to the nonlinearity of the collision term, the integro-differential equation \eqref{Boltzmann} is difficult to solve completely. Very rare analytical solutions to the nonlinear relativistic Boltzmann equation can be found \cite{Bazow:2015dha,Hu:2024ime}. Then one often resorts to  practical techniques to linearize the Boltzmann equation, among which Chapman-Enskog expansion and moment expansion are considered standard methods \cite{DeGroot:1980dk}. In the long-wavelength limit, where the mean free path $\ell_{\text{mfp}}$ of the system is significantly smaller than the typical length $L$ associated with system non-uniformity, hydrodynamics can be developed from kinetic theory via Knudsen number ($\frac{\ell_{\text{mfp}}}{L}$) expansion. 
 
 To proceed, the transport equation is linearized, resulting in the following equation
 \begin{align}
 \label{boltz}
D  f(x,p)+E_p^{-1}p ^{\langle\nu\rangle}\partial_\nu f(x,p)=-f_{0}(x,p)\mathcal{L}_0[\chi],
 \end{align}	
 where $D\equiv u\cdot\partial$, $E_p=u\cdot p$ and $p^{\langle\nu\rangle}\equiv \Delta^{\nu\rho}p_\rho$. We write the linearized collision operator as
 \begin{align}
 \label{cl}
 -\mathcal{L}_0[\chi]&\equiv E_p^{-1}\int  {\rm dP^\prime}  {\rm dP_1} {\rm dP_2}f_{0}(x,p^\prime)W_{p,p^\prime\to p_1,p_2} \big(\chi(x,p_1)+\chi(x,p_2)-\chi(x,p)-\chi(x,p^\prime)\big),
 \end{align}
 where an  expansion is  made according to $f=f_{0}(1+\chi)$ with a local equilibrium distribution defined as
 \begin{align}
 \label{feq}
 &f_{0}(x,p)=\exp[\xi-\beta\cdot p].
 \end{align}
  Here $\beta^\mu\equiv\frac{u^\mu}{T},\xi\equiv \frac{\mu}{T},\beta\equiv\frac{1}{T}$ with the temperature $T$,  and  the chemical potential $\mu$  associated with the conserved particle number. By this construction, the local equilibrium distribution ensures that the right-hand side (RHS) of the transport equations Eq.(\ref{Boltzmann}) and (\ref{boltz}) equals zero, 
  $C[f_{0}]=\mathcal{L}_0[f_{0}]=0$, due to the invariance of 1 and $p_{\mu}$ under collisions. Particularly, if the left-hand side (LHS) of Eq.(\ref{boltz}) is also zero i.e. $p\cdot \partial f_{0}=0$, this special state of local equilibrium is called global equilibrium represented  by $f_{eq}$, in which  we have
   \begin{align}
   \label{condition}
   &\partial_{(\mu}\beta_{\nu)}=0,\quad \xi=\text{const}.
   \end{align}
  It can be verified that $\mathcal{L}_0$ is  self-adjoint and positive semidefinite \cite{DeGroot:1980dk} interpreted as a linear operator in square-integrable Hilbert space
 \begin{align}
 &\int  {\rm dP} f_{0}(p)E_p\psi(p)\mathcal{L}_0\phi(p)= \int  {\rm dP}f_{0}(p)E_p \phi(p)\mathcal{L}_0\psi(p),\nn\\
 & \int  {\rm dP} f_{0}(p)E_p\psi(p)\mathcal{L}_0\psi(p) \geq 0.
 \end{align}
 Moreover, $\mathcal{L}_0$ respects the following property, 
 \begin{align}
 \label{zero}
 \mathcal{L}_0[\psi]=0,   \quad \psi=a+b\cdot p,
 \end{align}
 with $a$ and $b^\mu$ being the free parameters independent of $p^\mu$.  Note the collision invariants are the  zero modes in the eigenspectrum of $\mathcal{L}_0$.
 
  The structure of the linearized Boltzmann equation is reminiscent of the time-dependent Schrödinger equation. Likewise, a comprehensive understanding of the eigenspectrum of $\mathcal{L}_0$  is necessary for addressing the evolution equation. Therefore, solving the eigenspectrum of $\mathcal{L}_0$  is almost equivalent to resolving the linearized transport equation. To our knowledge, the first analytical eigenspectrum was obtained by C.S.Wang Chang and U.E.Uhlenbeck in the context of monatomic gases in the non-relativistic case (see chapter IV of \cite{deboer})\footnote{After finishing this manuscript,  we discovered a related work \cite{Denicol:2022bsq} recently, where the authors analytically determine all the eigenvalues and eigenfunctions of the linearized Boltzmann collision operator in massless scalar $\lambda\phi^4$ theory. }
  
If we opt to expand the distribution function around a global equilibrium configuration, $f=f_{eq}(1+\chi)$, Eq.(\ref{cl}) can be further simplified as 
 \begin{align}
 \label{boltz5}
 \partial_t \chi(x,p)+\bm{v}\cdot\nabla \chi(x,p)=-\mathcal{L}_0[\chi],
 \end{align} 
 where $\bm{v}\equiv \frac{\bm{p}}{p_0},\nabla_\alpha\equiv \Delta^\beta_\alpha\partial_\beta$. It should be noted that $f_{0}$ in Eq.(\ref{cl}) is replaced by $f_{eq}$  accordingly in this case.
Neglecting the dependence of $\chi$ on the spatial coordinate, one can reduce \eqref{boltz5} into a form of  pure momentum transport,
 \begin{align}
 \label{boltz1}
 \partial_t \chi(p,t)=-\mathcal{L}_0\chi(p,t),
 \end{align}
 which can be formally solved
 \begin{align}
 \label{chi}
 \chi(p,t)=e^{-\mathcal{L}_0t}\chi(p,0),
 \end{align}	 
 where $\mathcal{L}_0$ acts like the Liouville evolution operator. This equation  indicates that the deviation from equilibrium is exponentially decaying via particle collisions except for five zero modes. Hence, for  late-time dynamics, it is unnecessary to consider all eigenmodes, and a finite number suffices. 
 
  \section{ A model with a mutilated relaxation time}
  \label{relax}
 Solving the eigenspectrum of $\mathcal{L}_0$ or giving calculations with the full $\mathcal{L}_0$  is indeed a challenging task.	If we do not insist on solving  the linearized transport problem precisely,  we are inclined to approximate $\mathcal{L}_0$ with a simplified collision operator $\mathcal{L}$.  For example, the relaxation time approximation (RTA) \cite{1974Phy....74..466A}  proposed by Anderson and Witting  retains the smallest nonzero eigenvalue of $\mathcal{L}_0$ \footnote{It is  found that this approximation relies on the gapped eigenspectrum of $\mathcal{L}_0$, which holds only for hard interactions \cite{dud,dud1,Hu:2024tnn}.}. However, the RTA has a basic shortcoming:  one of the important properties of $\mathcal{L}_0$ --- the collision invariance ---  has not been retained. The lack of collision invariance arises from excluding zero eigenvalues in this approximation. Although according to  Eq.(\ref{chi}), these zero modes don't evolve with time, excluding them is problematic. To fix the issue of lacking  collision invariance, one has necessarily to impose the conservation laws manually on the macroscopic side. A more natural and straightforward approach to restore the collision invariance is to rewrite the RTA collision operator on the side of the  Boltzmann equation, as shall be given below.
  
 Following the principle outlined above, one can approximate the full linearized collision operator with a mutilated  operator (see chapter V of \cite{deboer}),
\begin{align}
\label{L}
-\mathcal{L}_0\simeq-\mathcal{L}= -\gamma_s+\gamma_s\sum_{n=1}^{5}|\lambda_n\rangle\langle \lambda_n|,
\end{align}
where $\gamma_s\sum_{n=1}^{5}|\lambda_n\rangle\langle \lambda_n|$ is the counter term introduced for recovering the collision invariance, see also \cite{Hu:2024tnn} for a new method to construct the mutilated operator by properly truncating $\mathcal{L}_0$ rather than adding counter terms. Here $|\lambda_n\rangle$ represents the orthonormal eigenfunctions of $\mathcal{L}_0$ and $\gamma_s$ is the smallest positive eigenvalue with the dimension of $[E]$. Now the RHS of Eq.(\ref{L}) explicitly retains the collision invariance of $\mathcal{L}_0$: $\mathcal{L}|\lambda_n\rangle=0, n=1,\cdots 5$. One can verify this is  a kind of RTA by expressing $\gamma_s$ as the relaxation time  $\tau_R\equiv\frac{1}{\gamma_s}$. Additionally, $\mathcal{L}|\lambda_n\rangle=\gamma_s |\lambda_n\rangle \cdots, n>5$. This is the origin of the term “mutilated": all  positive eigenvalues collapse into the smallest positive eigenvalue $\gamma_s$.  
  
An additional advantage over the traditional Anderson and Witting RTA (AW RTA) is that the relaxation time is permitted to have energy dependence.  In some special models,  a flexible parameterization for the relaxation time  is often employed \cite{Dusling:2009df,Dusling:2011fd,Kurkela:2017xis}
  \begin{equation}
  \label{tauR}
  \tau_R = (\beta E_p)^\alpha t_R,
  \end{equation}
  where $\alpha$ is an arbitrary constant, and $t_R$ has no energy dependence. The integration of diverse energy dependencies is thought to elucidate distinctive characteristics of bottom-up thermalization and shed light on the concealed elements contained in the full kinetic description \cite{Kurkela:2017xis}. The parameter $\alpha$ is contingent upon the specific dynamics involved and aligns with various physical contexts: $\alpha=0$ represents the conventional AW RTA framework, $t_R=\frac{1}{\gamma_s}$ \cite{1974Phy....74..466A}; $\alpha=0.38$ is posited to effectively mirror the kinetic descriptions within quantum chromodynamics (QCD) \cite{Rocha:2021zcw,Dusling:2009df,Dusling:2011fd}; whereas $\alpha=0.5$ serves as a good model for highly non-equilibrium scenarios, such as jet interactions, where $\tau_R$  corresponds to the well-known jet stopping time in this case \cite{Baier:1996kr,Baier:1996sk}. Above all, the energy-dependent RTA has found many motivated applications in various physical scenarios as an effective model. Nevertheless, one shall find that naively substituting Eq.(\ref{tauR}) into the RTA collision operator would cause an inconsistency.
  
 
  
 To show it, we  write the conservation equation for the energy-momentum tensor 
   \begin{align}
   T^{\mu\nu}\equiv\int  {\rm dP}f(x,p) p^\mu p^\nu .
   \end{align}
 With $\mathcal{L}_0\simeq \mathcal{L}_{\text{RTA}}=\frac{1}{\tau_R}$, we obtain
  \begin{align}
  \label{conserve}
  \partial_\mu T^{\mu\nu}=-\int  {\rm dP}\,p^\nu \frac{ u\cdot p}{\tau_R} f_0\chi.
  \end{align}
  This follows from the Boltzmann equation in the relaxation time approximation
  \begin{align}
  p\cdot \partial f(x,p)=-\frac{u\cdot p}{\tau_R} f_0(x,p)\chi(x,p).
  \end{align}

  The RHS of Eq.(\ref{conserve}) is not naturally zero unless we  impose the following condition
    \begin{align}
    \label{match0}
   \int  {\rm dP}\, p^\nu \frac{ u\cdot p}{\tau_R} f_0\chi=0.
    \end{align}
  When  the relaxation time $\tau_R$ is energy-independent as prescribed in the traditional RTA,  the Landau matching condition is recovered
  \begin{align}
    \label{match1}
    \int  {\rm dP} p^\nu u\cdot pf_{0}\chi=0.
  \end{align}
 However, if $\tau_R$ is  permitted to vary with energy, Eq.(\ref{match0}) is not consistent with the  conventional Landau matching condition Eq.(\ref{match1}). Notably, the mutilated RTA inherently includes all collision invariants, naturally leading to exact expressions of conservation laws upon integration independent of any matching condition or energy dependence of the relaxation time. 
  It is important to note that the mutilated RTA is extensively applied in nonrelativistic kinetic theory, and recent discussions have extended its application to relativistic contexts  \cite{Rocha:2021zcw,Hu:2022mvl,Hu:2022xjn}. 
  
  Below, we want to cast Eq.\eqref{L} into a less abstract form. Before that, we start  from  Eq.(\ref{boltz5}) within the mutilated relaxation time approximation given by Eq.(\ref{L}), and aim to find a solution in the form $\chi\sim\tilde{\chi}(k,p)e^{-ik\cdot x}$.  Then  a Fourier transformation leads us to \cite{degroot}
    \begin{align}
    \label{cf}
    &\tau \omega\tilde{\chi}+\hat{p}^\mu \kappa_\mu \tilde{\chi}=-iL\tilde{\chi},
    \end{align}
    where 
    \begin{align}
    \label{l11}
    L\tilde{\chi}\equiv \frac{1}{T^2}\mathcal{L}\tilde{\chi}=\gamma\tau\big(\tilde{\chi}-\sum_{n=1}^{5}(\tilde{\psi}_n,\tau\tilde{\chi})\tilde{\psi}_n\big),
    \end{align}
 with notations
 \begin{align}
 \label{pert}
 \tau\equiv \frac{p\cdot u}{T},\quad \omega\equiv \frac{u\cdot k}{T},\quad\hat{p}^\mu\equiv\frac{p^\mu}{T},\quad \kappa^\alpha\equiv\frac{\Delta^{\alpha\beta}k_\beta}{T},\quad \kappa\equiv\sqrt{-\kappa\cdot\kappa},\quad l^\alpha\equiv\frac{\kappa^\alpha}{\kappa}.
 \end{align}
 Here  $\gamma$ is the inverse of the relaxation time scaled by the temperature $T$, namely, $\gamma\equiv\frac{1}{\tau_R T} =\frac{\gamma_s}{T}$.
 The inner product is defined  as 
 \begin{align}
 \label{inner}
 (B,C)=\int  {\rm dP} f_{eq}(p)B(p)C(p),
 \end{align}
 where $f_{eq}(p)$, instead of $f_0(x,p)$, is utilized because we  now work with Eq.(\ref{boltz5}).
A direct correspondence exists  between  $\tilde{\psi}_n$, $(\tilde{\psi}_n,\tau \tilde{\psi}_m)$ and $|\lambda_n\rangle$, $\langle\lambda_n | \lambda_m\rangle$, implying that $(\tilde{\psi}_n,\tau \tilde{\psi}_m)=\langle\lambda_n | \lambda_m\rangle=\delta_{nm}$. By this construction, $L\tilde{\psi}_n=0, \, n=1,2\cdots 5$, therefore it can be shown that the macroscopic conservation laws are satisfied naturally.  
Below, we will employ the Gram-Schmidt orthogonalization method to construct $\tilde{\psi}_n$.


   All collision invariants constitute our data for constructing the linearized collision operator $L$
  \begin{align}
  \label{set0}
  &\tilde{\phi}_{1}= 1, \quad \tilde{\phi}_{2}=\beta u\cdot p\,,\quad \tilde{\phi}_{3}=\beta l\cdot p,\quad \tilde{\phi}_{4}=\beta j\cdot p, \quad \tilde{\phi}_{5}=\beta s\cdot p,
  \end{align}
  where  we introduce a triad $(l^\mu,j^\mu,s^\mu)$ to form an orthonormal set of unit vectors alongside  $u^\mu$, and
  two auxiliary unit vectors $j^\mu$ and $s^\mu$ are defined to satisfy
  \begin{align}
  &u\cdot l=u\cdot j=u\cdot s=l\cdot j=l\cdot s=j\cdot s=0,\nn\\
  &l^2=j^2=s^2=-1.
  \end{align}
  Thus $p^\mu$ can be formally expanded  as
  \begin{align}
  \label{basis0}
  &p^\mu=u\cdot p \,u^\mu-l\cdot p \,l^\mu-j\cdot p \,j^\mu-s\cdot p\, s^\mu.
  \end{align}
   In calculations using the rest frame $u^\mu=(1,0,0,0)$, this triad represents unit vectors aligned with $z,x,y$ directions, respectively.
 Using the established orthonormal relationship, the basis can be further refined through the Gram-Schmidt orthogonalization process
 \begin{align}
 \tilde{\varphi}_1=1, \quad \tilde{\varphi}_2=\beta (u\cdot p-\frac{e}{n}), \quad \tilde{\varphi}_3=\beta l\cdot p, \quad \tilde{\varphi}_4=\beta j\cdot p, \quad \tilde{\varphi}_5=\beta s\cdot p,
 \end{align}
 where $n, e$ denote the number  density and  energy density given in Appendix.\ref{int}.
 Note that $\tilde{\varphi}_i$ are orthogonal but not yet normalized. Normalization is achieved as follows
 \begin{align}
 \label{set}
 &\tilde{\psi}_{1}=\frac{\tilde{\varphi}_1}{\sqrt{V_{1,1}}}, \quad \tilde{\psi}_{2}=\frac{\tilde{\varphi}_2}{\sqrt{V_{2,2}}},\quad \tilde{\psi}_{3}=\frac{\tilde{\varphi}_3}{\sqrt{V_{3,3}}},\quad \tilde{\psi}_{4}=\frac{\tilde{\varphi}_4}{\sqrt{V_{4,4}}}, \quad \tilde{\psi}_{5}=\frac{\tilde{\varphi}_5}{\sqrt{V_{5,5}}}, 
 \end{align}
 where 
\begin{align}
\label{vij}
V_{i,j}\equiv \int  {\rm dP}f_{eq}(p)\tilde{\varphi}_{i} \tau \,\tilde{\varphi}_{j}.
\end{align}
 For simplicity in the remainder of this paper, we adopt the rest frame $u^\mu=(1,0,0,0)$.

\section{Energy-independent relaxation time and onset transition} 
\label{analysis}
In this section, we  assume that $\gamma$ is independent of energy, i.e., $\gamma=\frac{1}{\tau_R T} =\frac{\gamma_s}{T}$. We conduct a comprehensive normal mode analysis that extends beyond the long-wavelength approximation.  Our discussion is analogous to the calculation of retarded correlators presented in \cite{Romatschke:2015gic}, and we follow the same foundational assumptions, treating the particles as massless. This assumption facilitates the analytical determination or significant simplification of complicated momentum integrals. In certain physical contexts, such as the quark-gluon plasma generated in ultra-relativistic heavy-ion collisions, the particles in the medium are often considered massless. However, it is also possible to relax the ultra-relativistic assumption to account for massive transport.

 We first define the fluctuation amplitudes as
\begin{align}
\label{rhodef}
\rho_n(k)\equiv (\tilde{\psi}_n,\tau\tilde{\chi}(k,p)\,),
\end{align}
with a special case for $n=2$
\begin{align}
\label{rhodef1}
\rho_2(k)\equiv \frac{1}{\sqrt{V_{2,2}}}(\tau,\tau\tilde{\chi}(k,p)\,).
\end{align}
Note that  $\tilde{\chi}(k,p)$ represents the  deviation of the distribution from equilibrium, and thus $\rho_n$ can be interpreted as the fluctuation amplitudes of the conserved charge densities.

The Fourier transformed linearized kinetic equation Eq.(\ref{cf})  is then expressed as
\begin{align}
\label{lke}
&(c+\frac{\hat{p}}{\tau}\cdot l) \tilde{\chi}=\hat{\gamma}\sum_{n=1}^{5}(\tilde{\psi}_n,\tau\tilde{\chi})\tilde{\psi}_n=\hat{\gamma}\sum_{n=1}^{5}\rho_n\bar{\psi}_n, 
\end{align} 
where the new notations introduced are
\begin{align}
\label{cga}
c\equiv \frac{-i\omega+\gamma}{-i\kappa},\quad\,\hat{\gamma}\equiv i\frac{\gamma}{\kappa}.
\end{align}

We have reorganized the expansion basis so that $\bar{\psi}_1=\tilde{\psi}_1-\frac{\beta e \sqrt{V_{1,1}}}{nV_{2,2}}\tau+\frac{\beta^2 e^2 \sqrt{V_{1,1}}}{n^2V_{2,2}}$ and $\bar{\psi}_i=\tilde{\psi}_i$ for $i \neq 1$. This adjustment is due to the distinct definition of $\rho_2$, which is the fluctuation amplitude of the energy density. 
To get $\tilde{\chi}$,  the inverse of $(c+\frac{\hat{p}}{\tau}\cdot l)$ is required. Consequently, we must consider two distinct scenarios as outlined below.

\subsection{Collective modes}
Using spherical coordinates to represent the momentum $p$,  we express $\{\frac{\hat{p}}{\tau}\cdot l,\frac{\hat{p}}{\tau}\cdot j,\frac{\hat{p}}{\tau}\cdot s\}$  as $\{\cos\theta,\sin\theta\cos\phi,\sin\theta\sin\phi\}$. 
Provided that  the factor of $(c+\cos\theta)$ is not zero, then the reversion can be made safely
\begin{align}
\label{solution}
\tilde{\chi}=\frac{\hat{\gamma}\sum_{n=1}^{5}\rho_n\bar{\psi}_n}{c+\cos\theta}.
\end{align}

By substituting Eq.(\ref{solution}) into Eq.(\ref{rhodef}) and (\ref{rhodef1}), we derive a system of equations  for $\rho_n$
\begin{align}
\label{rho1}
\rho_1&=\sum_{i=1}^{5}\frac{\rho_i}{\sqrt{V_{1,1}}}\int  {\rm dP} f_{eq}(p) \frac{\hat{\gamma}\tau\bar{\psi}_i}{c+\cos\theta}=\hat{\gamma}(\coth ^{-1}c)\,\rho_1+\frac{3\hat{\gamma}\big(1-c\coth ^{-1}c\big)}{2}\rho_3,\\
\rho_2&=\sum_{i=1}^{5}\frac{\rho_i}{\sqrt{V_{2,2}}}\int  {\rm dP} f_{eq}(p) \frac{\hat{\gamma}\tau^2\bar{\psi}_i}{c+\cos\theta}=\hat{\gamma}(\coth ^{-1}c)\, \rho_2+2\sqrt{3}\hat{\gamma}\big(1-c\coth ^{-1}c\big)\rho_3,\\
\label{rho3}
\rho_3&=\sum_{i=1}^{5}\frac{\rho_i}{\sqrt{V_{3,3}}}\int  {\rm dP} f_{eq}(p) \frac{\hat{\gamma}\tau^2\cos\theta\bar{\psi}_i}{c+\cos\theta}=\frac{\sqrt{3}}{2}\hat{\gamma}\big(1-c\coth ^{-1}c\big)\rho_2-3\hat{\gamma}c\big(1-c\coth ^{-1}c\big)\rho_3,
\end{align} 
and
\begin{align}
\label{rho4}
\rho_4=\sum_{i=1}^{5}\frac{\rho_i}{\sqrt{V_{4,4}}}\int  {\rm dP} f_{eq}(p) \frac{\hat{\gamma}\tau^2\sin\theta\cos\phi\bar{\psi}_i}{c+\cos\theta}=\frac{3}{2}\hat{\gamma}\rho_4\big( (1-c^2)\coth^{-1}c+c\big),\\
\label{rho5}
\rho_5=\sum_{i=1}^{5}\frac{\rho_i}{\sqrt{V_{5,5}}}\int  {\rm dP} f_{eq}(p) \frac{\hat{\gamma}\tau^2\sin\theta\sin\phi\bar{\psi}_i}{c+\cos\theta}=\frac{3}{2}\hat{\gamma}\rho_5\big( (1-c^2)\coth^{-1}c+c\big),
\end{align} 
where the equations for $\rho_4$ and $\rho_5$ are degenerate. As a reminder, $\coth^{-1}c$  represents the inverse hyperbolic cotangent  of $c$.
It is evident from the preceding equations that $\rho_1$, $\rho_2$, and $\rho_3$ are interdependent. Consequently, Eqs.(\ref{rho1}) to (\ref{rho3})  form a system of homogeneous equations aimed at determining  $\rho_1$, $\rho_2$ and $\rho_3$. Given that these fluctuation amplitudes are well-defined with specific physical meaning in kinetic theory, their existence must be ensured, implying they must satisfy Eqs.(\ref{rho1}) to (\ref{rho3}). In essence, the secular equation for this system must be solvable
\begin{align}
\label{secular}
\Phi(\hat{\gamma},c)=\left(\hat{\gamma}  \coth ^{-1}c-1\right) \left(3 \hat{\gamma}  (c-\hat{\gamma} )+  (3 c\hat{\gamma} (\hat{\gamma} -c)-\hat{\gamma}) \coth ^{-1}c+1\right)=0.
\end{align}
Observe that $\Phi$ can be decomposed into a sound sector and a heat sector
\begin{align}
\label{secular1}
\Phi_1(\hat{\gamma},c)&= 3 \hat{\gamma}  (c-\hat{\gamma} )+  (3 c\hat{\gamma} (\hat{\gamma} -c)-\hat{\gamma}) \coth ^{-1}c+1=0,\\
\label{secular2}
\Phi_2(\hat{\gamma},c)&=\hat{\gamma}  \coth ^{-1}c-1=0.
\end{align}
The designation of these sectors is based on their behaviors at long wavelengths, which will be detailed subsequently.

The complexity in addressing the secular equations,  Eq.(\ref{secular1}) and (\ref{secular2}), arises from the presence of two free parameters, $c$ and $\hat{\gamma}$. Given that  $c$ is  a complex number, this effectively results in three independent parameters. Our current objective is not to solve for the fluctuation amplitudes but to ascertain the conditions under which solutions exist. This is intrinsically connected to the presence of collective modes at intermediate and large wavenumbers. Note as an aside, the existence and dispersion relations of collective hydrodynamic modes at long wavelengths are  well-known.

To that end, we apply the residue theorem, also known as the winding number theorem, which states that  the number of zeros of $\Phi$ within a region of the complex $c$-plane in which $\Phi$ is analytic, is equivalent  to the number of times the image  of  $\Phi$ encircles the origin in the complex plane. During this analysis,  $\hat{\gamma}$ is held fixed. 

Let us  begin by examining $\Phi_1$ first. We opt to fix $\hat{\gamma}$ and vary $c$ to trace the path of $\Phi_1$.
Initially, the asymptotic behavior as $c\rightarrow \infty$ while $\hat{\gamma}$ remains finite is readily available
\begin{align}
\Phi_1(\hat{\gamma},c\rightarrow \infty)=1,
\end{align}
which simplifies the visualization of the path.  As $c$ runs along the upper real axis from $c=-\infty+i0^{+}$ to $c=\infty+i0^{+}$ and encompasses the large semicircle,  our task is to sketch the trajectory of $\Phi$ and determine how many times it encircles the origin. A representative diagram is provided for  $\hat{\gamma}=5i$.
\begin{figure}
	\centering
	{\includegraphics[width=0.45\linewidth]{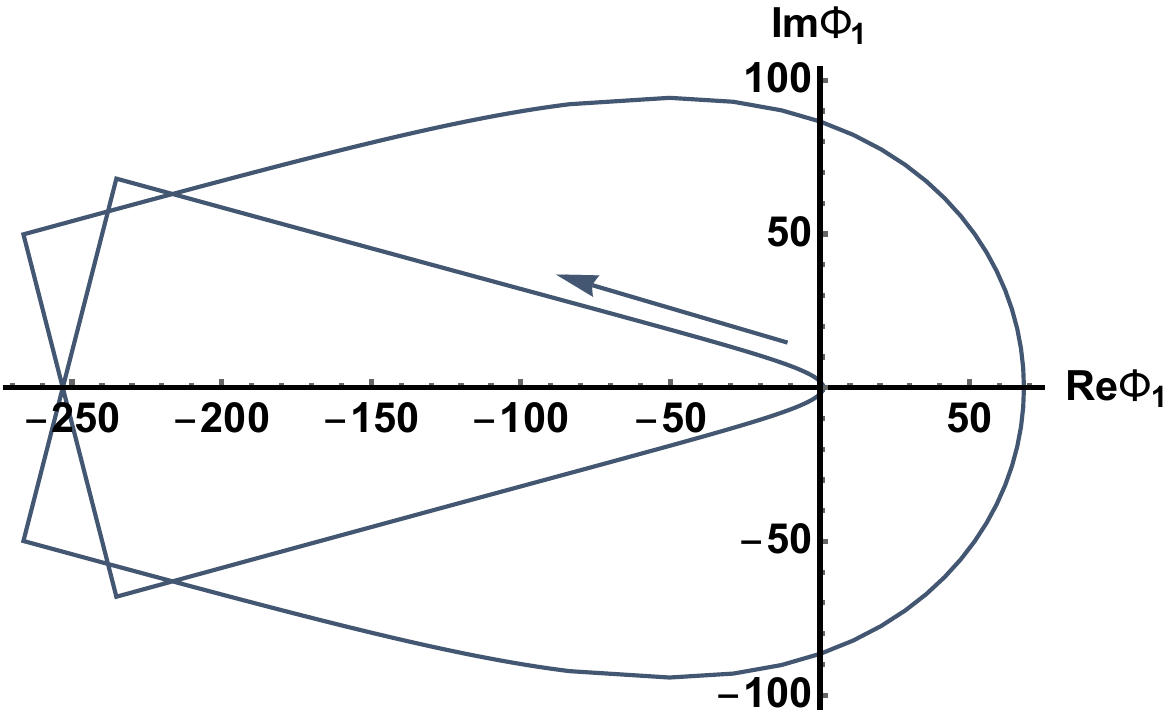}}
	{\includegraphics[width=0.42\linewidth]{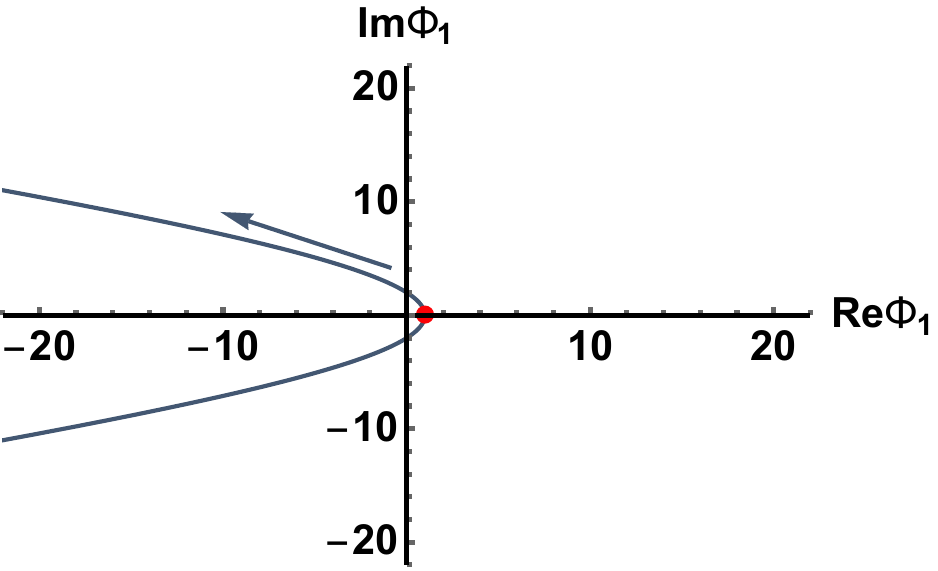}}
	\caption{The typical trajectories of $\Phi_1(5i,c)$. The right panel is an enlarged view of the  left panel around the origin. The curve starts from $\Phi_1(5i,-\infty+i0^{+})=1$ (red dot) and  encircles the origin twice along the direction of the arrow, while $c$ goes from $-\infty+i0^{+}$ to $\infty+i0^{+}$ and runs across the large semicircle back to $-\infty+i0^{+}$. }
	\label{traject}
\end{figure}

As depicted in Fig.\ref{traject}, when $c$ runs along the boundary of the upper complex plane, the trajectory of $\Phi_1$ encircles the origin twice, corresponding to the two collective modes recognized in relativistic hydrodynamics:  two sound modes. One might wonder about the broken line  appearing in the graph; this is due to the presence of branch points for $\Phi_1$ or $\coth^{-1}c$ at $c=\pm 1$. Enhancing the density of data points around these branch points would result in a smoother curve, but such efforts are unnecessary because a detailed analysis of $\Phi_1$'s behavior  near $c=\pm 1$ confirms that the overall winding number remains unaffected. It is important to note that when applying the residue theorem to the lower half-plane of $c$, no zeros are encountered.

\begin{figure}
	\centering
	{\includegraphics[width=0.45\linewidth]{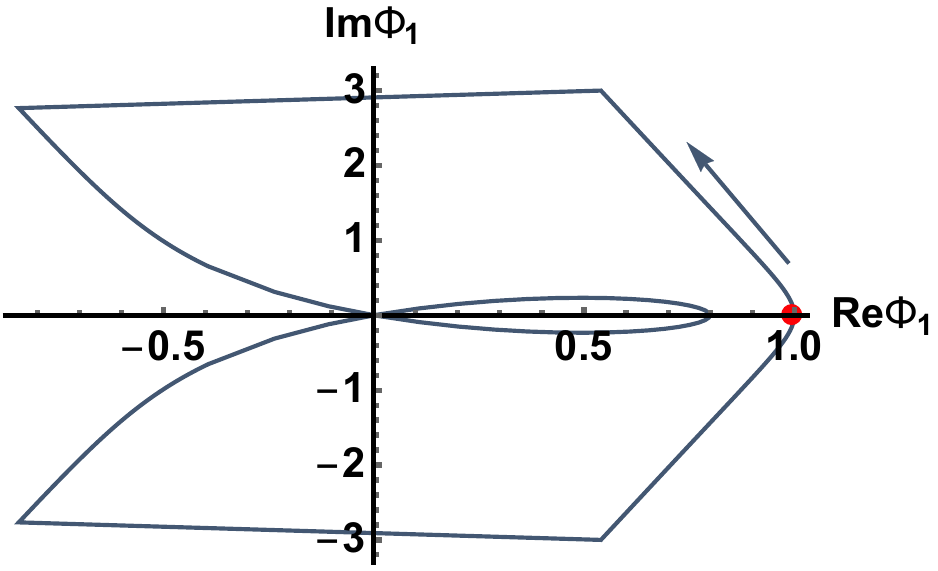}}
	{\includegraphics[width=0.41\linewidth]{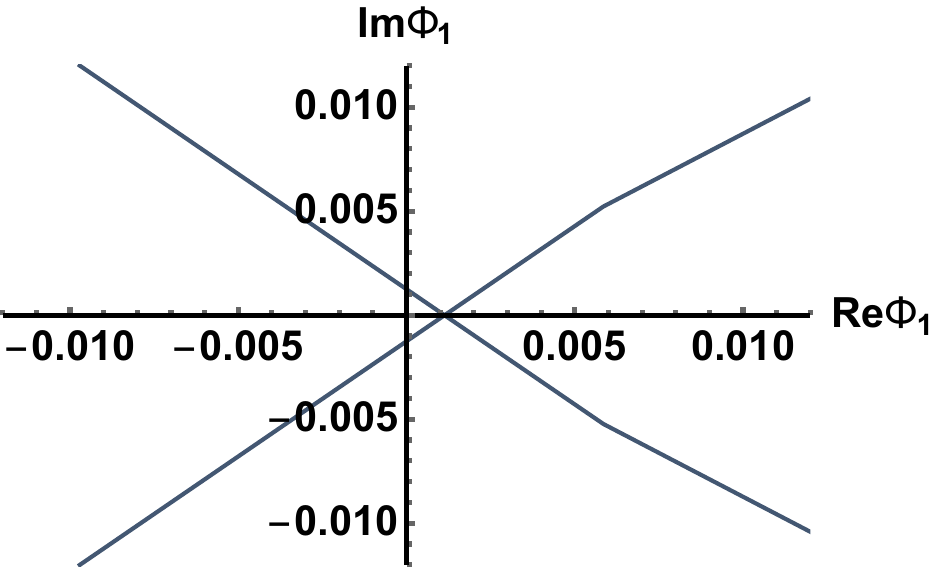}}	
	\caption{The typical trajectories of $\Phi_1(0.2206i,c)$. The right panel is an enlarged view of the  left panel around the origin. Starting from $\Phi_1(0.2206i,-\infty+i0^{+})=1$ (red dot), the curve runs along the direction of the arrow, while $c$ goes from $-\infty+i0^{+}$ to $\infty+i0^{+}$ and runs across the large semicircle back to $-\infty+i0^{+}$. The origin remains unenclosed.}
	\label{none}
\end{figure}

Next, we aim to investigate the behavior of these collective modes at short wavelengths. Changes occur gradually as we decrease the ratio $\frac{\gamma}{\kappa}$ or increase the value of $\kappa$. A transition occurs at the critical ratio $\frac{\gamma}{\kappa}=0.2208$, below which the two  sound modes cease to exist, as illustrated in Fig.\ref{none}. It is worth noting that a study as detailed in \cite{Romatschke:2015gic},  has previously identified this phenomenon, referring to it as the hydrodynamic onset transition. The threshold value of  $\gamma$ is determined to be  $\frac{\gamma}{\kappa}=0.2207$,  which is in close proximity to our  result. Applying a similar analysis to $\Phi_2$ reveals the same transitional behavior for the heat mode, with a critical value at $\frac{\gamma}{\kappa}\simeq\frac{2}{\pi}$. Below this value, the heat mode ceases to exist. Frankly, our approach, which employs the winding theorem to determine the critical ratio $\frac{\gamma}{\kappa}$,  yields an approximate numerical  value. By comparing our results with the analytical calculations given in \cite{Romatschke:2015gic}, we confirm that the numerical value is  in exact correspondence with $\frac{2}{\pi}$. 

Eventually, we turn our attention to the shear-channel modes. The existence condition, as implied by \eqref{rho4}, is given by
\begin{align}
\label{phi3}
\Phi_3(\hat{\gamma},c)=\frac{3}{2}\hat{\gamma}\big( (1-c^2)\coth^{-1}c+c\big)-1=0.
\end{align}

By applying similar procedures as before, we determine the critical value for $\hat{\gamma}$, which approximates to $\frac{\gamma}{\kappa}\simeq\frac{4}{3\pi}$.  As with previous cases, we note that this is an approximate value. For illustration,  we introduce an alternative analytical method to re-evaluate the critical value.
 First, we postulate a purely decaying dispersion relation for the shear-channel modes, given by  $\omega=-i\gamma(1+\alpha(\kappa/\gamma)\,)$  where $\alpha(\kappa/\gamma)$ real-valued function of  $\kappa/\gamma$. Then, according to the definition of  $c$, $c$ should be purely imaginary 
 $c=-i\frac{\alpha(\kappa/\gamma)}{\kappa/\gamma}$.
Thus we reformulate Eq.(\ref{phi3}) as 
\begin{align}
\label{shear}
(1+(\frac{\alpha}{\kappa/\gamma})^2)\cot^{-1}[\frac{\alpha}{\kappa/\gamma}]-\frac{\alpha}{\kappa/\gamma}=-\frac{2}{3}\frac{\kappa}{\gamma},
\end{align}
utilizing the relationship between $\coth^{-1}c$ and $\cot^{-1}c$.
Fortunately, $\frac{\alpha}{\kappa/\gamma}$ and $\frac{\kappa}{\gamma}$ are decoupled, simplifying the analysis. It is physically reasonable to assume that $\frac{\kappa}{\gamma}$ is positive. The critical value for the shear-channel transition can thus be ascertained, yielding $\frac{\gamma}{\kappa}=\frac{4}{3\pi}$. 

With all the necessary components in place, the critical values of $\frac{\gamma}{\kappa}$ for the various collective modes have been  determined numerically and are presented in  Table.~\ref{table1}. 
\begin{table}[!hbt]
	\begin{tabular}{c|c|c|c}
		\hline\hline
		&\quad Sound \quad
		&\quad shear \quad
		& \quad heat  \quad		
		\\
		\hline
		winding theorem
		& 0.2208 
		& $\simeq\frac{4}{3\pi}$
		& $\simeq\frac{2}{\pi}$	 \\	
      \hline
     empirical analytical calculation
     & 0.2207 
     & $\frac{4}{3\pi}$
     & $\frac{2}{\pi}$	 \\		
		\hline\hline
	\end{tabular}
	\caption{The critical values $\frac{\gamma}{\kappa}$ for various hydrodynamic modes in relativistic hydrodynamics. The numerical calculations are conducted with a precision such that   "$\simeq$" is accurate to within an order of $10^{-7}$.} 
	\label{table1}
\end{table}

Typically, one can numerically work out the specific dispersion relations for these collective modes. In some limits, even analytical results can be sought. To elucidate this, we  demonstrate how to identify their hydrodynamic dispersion relations through a hydrodynamic expansion, using the heat channel as an example. Recalling the definition of $c$, a hydrodynamic expansion around $\kappa=0$ would send $c$ to infinity. The asymptotic behavior of $\coth^{-1}c$ is given by
	\begin{align}
	\coth^{-1}c=\frac{1}{2}\ln \frac{c+1}{c-1}=\sum_{n=0}^{\infty}\frac{1}{2n+1}\frac{1}{c^{2n+1}},
	\end{align}
	which can be approximated by $\frac{1}{c}+\frac{1}{3c^3}+\mathcal{O}(\frac{1}{c^5})$ for large $|c|$. Substituting this into $\Phi_2$, we obtain
	\begin{align}
	\Phi_2(\hat{\gamma},c)=\hat{\gamma}  (\frac{1}{c}+\frac{1}{3c^3})-1=0,
	\end{align}
	and upon inserting the expression $c\equiv \frac{\omega+i\gamma}{\kappa}$ into $\Phi_2$, 
	we reach $\omega=-i\frac{1}{3\gamma}\kappa^2+\mathcal{O}(\kappa^4)$, reproducing the well-known hydrodynamic dispersion relation for the heat mode. 
	
	By applying similar calculation procedures, the dispersion relations for other hydrodynamic modes can be determined and are summarized as follows
	\begin{align}
	\label{sound}
	\omega&=\pm c_s\kappa-\frac{2i}{15\gamma}\kappa^2,\quad\quad \text{sound modes}\\
	\omega&=-\frac{i}{3\gamma}\kappa^2,\quad\quad \text{heat mode}\\
	\label{shear1}
	\omega&=-\frac{i}{5\gamma}\kappa^2,\quad\quad \text{shear modes, twofold degeneracy}
	\end{align}
	where the sound speed $c_s$ approaches $\frac{1}{3}$ in the conformal limit. These are consistent with the number of zeros predicted by the winding theorem.
	Thus, we label the various channels according to their hydrodynamic correspondence. These kinetic modes emerge from the zero eigenvalues of the collision operator, reflecting the organized collective motion of particles as orchestrated by collisions.


Before ending this section, several comments are followed in order.
 \begin{itemize}
 	\item
 	The normal mode analysis detailed here extends the conventional hydrodynamic mode analysis \cite{DeGroot:1980dk,Hu:2022azy,Perna:2021osw,Hu:2022lpi} to transient and short-wavelength regimes, encompassing a complete order calculation in wavenumber $k$. While the model, which simplifies the linearized collision kernel to a single relaxation time, may not provide a fully quantitative description, it is invaluable for examining the analytical structure of hydrodynamic behaviors. 
 	 Crucially, the counter terms in Eq.(\ref{L}) are essential,  without which Eqs.(\ref{rho1}) to (\ref{rho5}) would not be achievable, highlighting a distinct advantage of the novel RTA over the traditional one.
 	\item
 	In pursuit to uncover the normal modes embedded in Eq.(\ref{secular}) and (\ref{phi3}),  and delineate the thresholds of onset transitions, our approach avoids making any presuppositions. This contrasts with the empirical methods used in \cite{Romatschke:2015gic}, where the dispersion relations of the normal modes were first assumed. Herein, it is preferable to employ the residue theorem. Both methods yield nearly identical results, serving as a consistent cross-check.
 	\item
 It is important to acknowledge the omission of sound-heat modes coupling  in this analysis. Although sound and heat channels appear to be interlinked, they essentially decouple as demonstrated in Eq.(\ref{secular1}) and (\ref{secular2}) 
 However, in the long-wavelength limit, the secular equations for determining the dispersion relations of the sound and heat modes are indeed coupled, known as a common feature in hydrodynamic calculations \cite{Hu:2022azy}. To clarify a potentially confusing aspect, we remind that two approximations are applied in the practical calculations: first, that particles are considered massless, and second, that the relaxation time is assumed to be independent of energy. If  either of these assumptions is relaxed, the resulting coefficient matrix in Eqs.(\ref{rho1}) to (\ref{rho3}) would generally contain nonzero elements. As an illustrative example, the  calculations in the non-relativistic limit detailed on page 207 of Chapter V  in \cite{deboer} demonstrate a matrix with no zero elements. 
 		
 The normal mode analysis presented in this script naturally incorporates sound-heat modes coupling if we relax either of the two approximations. In the context of retarded correlators, accounting for these couplings implies the inclusion of cross-correlation functions such as $G^R_{ne}, G^R_{en}$, necessitating calculations of $\frac{\delta T^{\mu\nu}}{\delta A^\alpha}$ and $\frac{\delta J^\mu}{\delta g^{\alpha\beta}}$ (refer to  \cite{Romatschke:2015gic,Bajec:2024jez} for definitions and notations. Ref.\cite{Bajec:2024jez} recently gives the calculations of those cross-correlators, and thus closes this gap). In this sense, the present model has access to a comprehensive  analysis with minimal adjustments. We note that these two primary  approximations, considering particles as massless and assuming an energy-independent relaxation time, are also used in \cite{Romatschke:2015gic}. Therefore, the omission of these effects should also be expected therein.
    \end{itemize}	
  
  \subsection{Non Collective modes}\label{noncoll}
  When the factor of $(c+\cos\theta)$  equals zero, it implies a specific relationship between the real and imaginary parts of  $\omega$,  as follows
  \begin{align}
  \label{condition1}
 \Re \omega\in [-\kappa,\kappa],\quad i\Im\omega=-i\beta\gamma_s.
  \end{align}
 This condition corresponds to the branch cut of the $\Phi$ functions discussed in the previous section. From
  Eq.(\ref{lke}), we obtain
  \begin{align}
  \label{nonhy}
  \tilde{\chi}=A(\omega,\kappa)\delta(c+\cos\theta)+\hat{\gamma}\sum_{n=1}^{5}\rho_n\bar{\psi}_n\text{P}\frac{1}{c+\cos\theta},
  \end{align}
where $\text{P}$ denotes the principal value, and we employed an identity $x\delta(x)=0$, which gives rise to the first  term on the RHS of Eq.(\ref{nonhy}). Here $A$ is an undetermined factor but its specific form does not affect the subsequent discussion.  For  clarity, our discussion is limited to sound-heat channels. Substituting Eq.(\ref{nonhy}) into the definition of $\rho_i$ yields a system of linear inhomogeneous equations
  \begin{align}
  \label{rho2}
  \rho_1&=\frac{A}{2}+\hat{\gamma}(\coth ^{-1}c)\rho_1+\frac{3\hat{\gamma}\big(1-c\coth ^{-1}c\big)}{2}\rho_3,\\
  \rho_2
  &=\frac{\sqrt{3}A}{2}+\hat{\gamma}(\coth ^{-1}c)\rho_2+2\sqrt{3}\hat{\gamma}\big(1-c\coth ^{-1}c\big)\rho_3,\\
  \rho_3
  &=-\frac{3Ac}{4}+\frac{\sqrt{3}}{2}\hat{\gamma}\big(1-c\coth ^{-1}c\big)\rho_2-3\hat{\gamma}c\big(1-c\coth ^{-1}c\big)\rho_3.
  \end{align} 
 Here $c$ is taken to be $\frac{\Re \omega}{\kappa}$. With straightforward algebraic manipulation, these fluctuation amplitudes can be solved
 \begin{align}
 \rho_1&=\frac{A \left(\hat{\gamma} \coth ^{-1}c \left(-9 \hat{\gamma}  c^2\coth ^{-1}c+3 c (5 c+\hat{\gamma} )+8\right)+6 \hat{\gamma} ^2-15 c \hat{\gamma} -8\right)}{16 \left(\hat{\gamma} \coth ^{-1}c-1\right) \left(3 \hat{\gamma}  (c-\hat{\gamma} )+\hat{\gamma}  (3 c (\hat{\gamma} -c)-1)\coth ^{-1}c+1\right)},\\
 \rho_2&=\frac{\sqrt{3} A \left(3 \hat{\gamma}  c^2\coth ^{-1}c-3 \hat{\gamma}  c-2\right)}{4 \left(3 \hat{\gamma}  (\hat{\gamma} -c)+\hat{\gamma}  (3 c (c-\hat{\gamma} )+1)\coth ^{-1}c-1\right)},\\
 \rho_3&=\frac{3 A \left(\hat{\gamma}  c\coth ^{-1}c+c-2 \hat{\gamma} \right)}{8 \left(3 \hat{\gamma}  (\hat{\gamma} -c)+\hat{\gamma}  (3 c (c-\hat{\gamma} )+1)\coth ^{-1}c-1\right)}.
 \end{align}
 Substituting the obtained $\rho_i$ back into Eq.(\ref{nonhy}) will give us an explicit solution to the normal mode problem. In this scenario, there is no functional relationship, or dispersion law,  between $\omega$ and $\kappa$ as a condition for the existence of collective modes. The delta function in  Eq.(\ref{nonhy}) signifies that these modes are merely single-particle modes associated with a continuum single-particle spectrum. In the free-streaming dominated region, we can neglect the collision effects in the kinetic equation, and obtain
   \begin{align}
   \label{nonhy1}
   \tilde{\chi}\sim A(\omega,\kappa)\delta(c+\cos\theta).
   \end{align}
 The initial fluctuation is carried away by disordered particles, indicating the absence of discrete kinetic modes. In this case, all continuum modes damp with a characteristic relaxation time $1/(\gamma T)$, as per Eq.(\ref{condition1}).
\subsection{A  complete description}
In the preceding subsections, we discussed the normal mode solutions within the linearized kinetic theory. These normal modes can be formally categorized into two groups. The first group comprises discrete kinetic modes with well-defined dispersion relations, which transition into hydrodynamic modes in the long-wavelength limit as exhibited in Eqs.(\ref{sound}) to (\ref{shear1}). The other group consists of the non-hydrodynamic modes associated with  the branch cut of the retarded correlation function presented in \cite{Romatschke:2015gic,Bajec:2024jez}. In our formalism, this information on the non-hydrodynamic modes can also be recovered, and the non-collective modes described in \ref{noncoll} exactly fall within this category. 

Recall that the  Eqs.(\ref{secular1}), (\ref{secular2}) and (\ref{phi3}) correspond to the denominators of the retarded correlators as formulated in \cite{Romatschke:2015gic}, and the zeros of these equations signify the hydrodynamic poles in the corresponding retarded correlators. On the other hand, the numerical analysis using the winding theorem  only confirms the existence of five collective modes. The additional non-hydrodynamic modes  correspond to the scenarios beyond the winding theorem's scope (note when drawing Figs.\ref{traject} and \ref{none}, we never touch the region of $c\in [-1,1]$). When $c+\cos\theta=0$, or equivalently when $c$ ranges from $-1$ to $1$ on the real axis, the discussion provided in \ref{noncoll} offers a complementary description of  the non-hydrodynamic modes.

  \begin{figure}
  	\centering
  	{\includegraphics[width=0.5\linewidth]{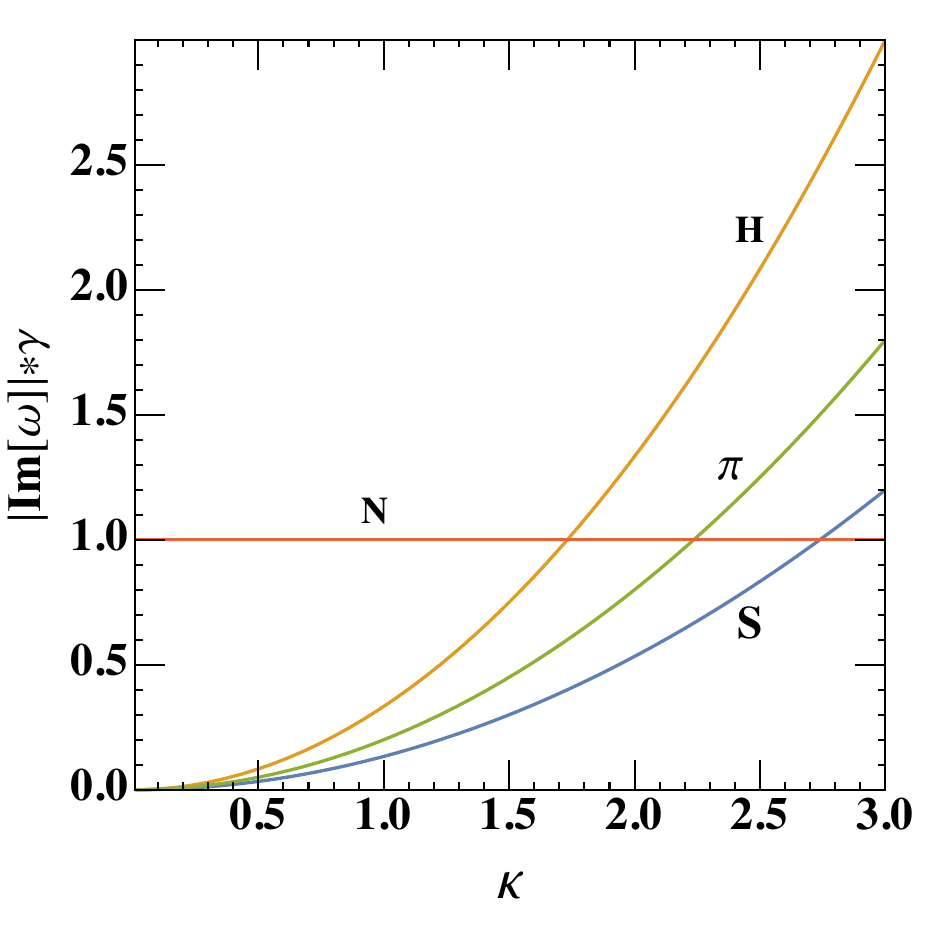}}
  	\caption{The imaginary parts of hydrodynamic modes as a function of $\kappa$ with $\gamma$ taken to be 1. S, H, $\pi$ and N denote sound modes, heat mode, shear modes and non-hydrodynamic modes respectively.}
  	\label{life}
  \end{figure}
  
     In conclusion, these non-hydrodynamic modes are essentially non-collective single-particle modes, associated with the branch cut of retarded correlators. 
   It is worth noting that the typical relaxation times for hydrodynamic and non-hydrodynamic modes are distinctly separated. The  momentum-independent relaxation time introduces a new scale or gap $\gamma$, as illustrated by the red solid line in Fig.\ref{life}.  Only low-energy effective degrees of freedom below this gap significantly contribute to late-time dynamics, thereby emphasizing that  hydrodynamics serves as a low-energy effective theory in the long-wavelength limit. The sharpness of onset transitions is attributed to the energy independence of the gap. In the following section, we will explore the interplay between collective hydrodynamic modes and non-collective non-hydrodynamic modes in the case of an energy-dependent relaxation time.  We will find that  the unique leading role of hydrodynamic modes in late-time evolution would be challenged by non-hydrodynamic modes.
   	
 
 

\section{Energy-dependent relaxation time and dehydrodynamization} 
\label{dep}
As outlined  in Sec.\ref{relax}, the mutilated RTA model inherently accommodates an energy-dependent relaxation time, which is an effective model in some special cases. In this section, we focus on examining the impact of the energy dependence on the normal mode analysis previously performed.

Given that $\gamma=\frac{1}{\tau_R T} =\frac{\gamma_s}{\tau T}$ where we define $\tau_R$ as given in Eq.(\ref{tauR}) with $\alpha=1$, and $\gamma_s$ is still interpreted as  the smallest nonzero eigenvalue  of $\mathcal{L}_0$, a minor adjustment is required in the definitions  of $c$ and $\hat{\gamma}$ appearing in Eq.(\ref{cga})
\begin{align}
c\equiv \frac{-i\omega+\beta\gamma_s/\tau}{-i\kappa},\quad\,\hat{\gamma}\equiv i\frac{\beta\gamma_s}{\kappa\tau},
\end{align}
 while other  expressions remain intact. 
 
 To proceed, we revisit Eq.(\ref{solution}). It is essential to invert  Eq.(\ref{lke}) to derive the formal expression for the deviation function $\tilde{\chi}$ before identifying collective kinetic modes. Caution must be taken in this inversion process, as the LHS of Eq.(\ref{lke}), containing $c$,  is dependent on $\tau$ which can range from zero to positive infinity. Consequently,  $(c+\cos\theta)$ could potentially be zero, provided that
   \begin{align}
   \label{condition3}
   \Re \omega\in [-\kappa,\kappa],\quad i\Im\omega=-i\beta\gamma_s/\tau.
   \end{align}
Thus, this factor can not be naively divided.
 
 For a given $\tau$, the classification criteria between collective modes and non-collective single-particle modes remain effective, with the branch cut shifting from Eq.(\ref{condition1}) to Eq.(\ref{condition3}). However,  as indicated in Eqs.(\ref{rho1}) to (\ref{rho5}),   $\tau$ is not a fixed external variable but an intermediate variable that will be integrated out.  Varying $\tau$ spanning from zero to infinity results in  a spectrum of branch cut lines.
 
 The integration over $\tau$ suggests a complex structure: rather than a single branch cut line, we have a strip composed of a series of branch cuts. This strip occupies the entire lower complex plane of $\omega$, with the real part of $\omega$ ranging from $-\kappa$ to $\kappa$. Besides, the integration over $\tau$  renders the division by $(c+\cos\theta)$ problematic, complicating the separation of collective modes from non-collective single-particle modes. This scenario can be visualized in Fig.\ref{life} : when there is only a single branch cut line, we can identify hydrodynamic modes below the branch line as low-energy effective degrees of freedom. However, when we translate the non-hydrodynamic branch line up and down to fill this rectangular half-plane, all hydrodynamic modes become embedded within this strip structure. Notably, the most energetic single-particle modes, which have a smaller absolute imaginary part, are intertwined with hydrodynamic modes, eliminating the gap below which hydrodynamic modes are the only low-energy degrees of freedom.
 
 To further elucidate this mathematically, the complicated integral in Eqs.(\ref{rho1}) to (\ref{rho5}) can be reduced to an expression including the following integral
 \begin{align}
I=\Big(-\partial_a\int_{-1}^{1}dx\int_0^\infty d\tau\frac{\exp(-a\tau)}{(\omega+\kappa x)\tau+i}\Big)_{a=1}.
 \end{align}
 This aligns with an integral formula from a related study, the details of which can be found in \cite{Kurkela:2017xis}. We opt to calculate it explicitly but only focus on the non-analytical structure, disregarding less significant factors.
 After performing the $\tau$ integration and applying integration by parts, the integral simplifies to
 \begin{align}
 \label{I}
 I\sim 2\pi i\exp(-1/\Im \omega)\theta(-\Im\omega)\theta(\kappa^2-\Re\omega^2),
 \end{align}
 where the nonanalytic strip structure is clearly recovered. This structure arises from the  $x$-integration crossing the branch cut of incomplete gamma function $\Gamma[0,\frac{ia}{\omega+x}]$, which emerges after the $\tau$ integration. In summary, both Eq.(\ref{condition3}) and (\ref{I}) indicate a similar nonanalytic strip structure for nonhydrodynamic modes.

 Revisiting the concept of  onset transitions, as discussed in the previous section, is quite enlightening. When the relaxation time is energy-independent, it introduces a unique gap, illustrated by the red solid line in Fig.\ref{life}. This gap delineates a window in the long-wavelength (low $\kappa$) region where only hydrodynamic modes are present. However, this window narrows significantly when considering an energy-dependent relaxation time. In such cases, regardless of how small  $\kappa$ is,  there will always be a series of  branch cut lines intersecting with or beneath the frequencies of these low $\kappa$ hydrodynamic modes. The omnipresence of non-hydrodynamic modes means that those with large $\tau$ could contribute significantly, or even more so, to the late-time dynamic evolution than hydrodynamic modes. This phenomenon is referred to as  “dehydrodynamization” proposed in \cite{Kurkela:2017xis}.
 

 
 However,  the above discussions are far from enough because the Boltzmann equation is known for its intricate eigenvalue spectrum, which includes both isolated and continuous eigenvalues. The characteristics of this spectrum are highly dependent on the specific interparticle interactions. Studies, such as those found in \cite{libo,dud}, have shown that for hard potentials, the continuous spectrum ranges from a finite value to infinity, which allows for practical approximations to represent the continuum with a single eigenvalue. In contrast, soft interactions yield a continuous spectrum that starts from zero. Therefore, to fully understand the complex eigenvalue spectrum, it is necessary to specify the interactions and conduct case-by-case calculations, as demonstrated in \cite{Moore:2018mma,Ochsenfeld:2023wxz,Denicol:2022bsq}.
 


 \section{Summary and outlook}
 \label{su}
 In this paper, we perform a comprehensive normal mode analysis based on the Boltzmann equation in the mutilated relaxation time approximation. Our linearized kinetic description can extend the scope of traditional hydrodynamic mode analysis to the intermediate and short-wavelength regions, providing a natural categorization of kinetic modes into collective modes and non-collective single-particle excitations.
 
 By utilizing an energy-independent relaxation time, we reproduce the anticipated behavior of hydrodynamic onset transitions, as observed in \cite{Romatschke:2015gic,Bajec:2024jez}. Given the assumptions of massless particles and energy-independent relaxation time, the expected sound-heat channel coupling is notably absent. In addition, in this scenario there is a gap below which hydrodynamic modes are the only low-energy effective degrees of freedom. However, the introduction of an energy-dependent relaxation time complicates this classification, as hydrodynamic modes are enveloped by a multitude of branch cuts, indicating a blend with non-hydrodynamic modes. A comprehensive account of late-time dynamic evolution must consider both collective and non-collective modes. It is worth noting that the second-order hydrodynamic theory by Israel and Stewart \cite{Israel:1979wp} encapsulates both two kinds of modes, potentially broadening its applicability beyond that of the first-order hydrodynamics.
 
We envision several potential extensions for this research. 
 The current approximation, which simplifies the linearized collision operator to a single relaxation term, inevitably neglects certain nuances. It would be beneficial to perform a similar normal mode analysis using the complete linearized collision operator to ascertain whether any critical elements are omitted. Lastly, numerous theoretical investigations into the behavior of hydrodynamic attractors are predicated on linearized kinetic models or more rudimentary frameworks like the RTA, frequently overlooking the transport equation's inherent nonlinearity. Delving into the profound impact of nonlinearity on the equilibration process, as exemplified by kinetic equations including the Boltzmann equation, represents a compelling area of exploration.

\section*{Acknowledgments}
J.Hu is grateful to Shuzhe Shi for reading this manuscript and helpful comments.
\begin{appendix}

\section{Thermodynamic  Integral} \label{int}
 We focus on the following integral of interest
\begin{align}
\label{inq}
I^{(r)}_{\alpha_1\cdots\alpha_n}&\equiv \int {\rm dP}\,f_{eq}(p)p_{\alpha_1}p_{\alpha_2}\cdots p_{\alpha_n}\frac{1}{(u\cdot p)^r}\nn\\
&=I^{(r)}_{n0}u_{\alpha_1}\cdots u_{\alpha_n}+I^{(r)}_{n1}(\Delta_{\alpha_1\alpha_2}u_{\alpha_3\cdots\alpha_n}+\text{permutations})+\cdots,
\end{align}
where  $\int {\rm dP}\equiv \frac{2}{(2\pi)^3}\int d^4p\, \theta(p^0)\delta(p^2 - m^2)$, and the formal expression in the second line  arises from the analysis of Lorentz covariance. By employing projection operators $u^\alpha$ and $\Delta^{\alpha\beta}$, the scalar coefficients in terms of thermodynamic integrals can be expressed as 
\begin{eqnarray}
\label{Inq}
I^{(r)}_{nq} &\equiv& \frac{2}{(2q+1)!!} \int  {\rm dP}\,f_{eq}(p)(u\cdot p)^{n-2q-r} (\Delta_{\alpha\beta} p^{\alpha} p^{\beta})^q ,
\end{eqnarray}
with $K_n(z)$  denoting the modified Bessel functions of the second kind defined as
\begin{eqnarray}
K_n(z) &\equiv& \int_0^{\infty} \mathrm{d}x\, \cosh(nx)\, e^{- z \cosh x}.
\end{eqnarray}
It is straightforward to show that $I^{(-r)}_{nq}=I^{(0)}_{n+r,q}$ for non-negative integer values of $r$. Since $(u\cdot p)^r$ does not influence the Lorentz tensor structure formed by $u^\mu$ and $\Delta^{\mu\nu}$, $r$ can take negative or fractional values. Specially, we have $I^{(0)}_{10}=n$, $I^{(0)}_{20}=e,\, I^{(0)}_{21}=-P$ , $I^{(0)}_{31}=-hT$ and 
\begin{align}
I^{(0)}_{30}(z)=\frac{T^5z^5e^{\xi}}{32\pi^2}\big(K_5(z)+K_3(z)-2K_1(z)\big),
\end{align}
where $z\equiv\frac{m}{T}$, $n, e, P, h$ are the number  density, energy density, static pressure, and enthalpy density respectively.


\section{Normalized factors} \label{int1}

The integrals below are instrumental in the Gram-Schmidt orthogonalization process and in determining the eigenfunctions with zero eigenvalues of the linearized collision operator,
\begin{align}
V_{i,j}=\int  {\rm dP}f_{eq}(p)\tilde{\varphi}_i \tau\,\tilde{\varphi}_j,
\end{align}
which follows from Eqs.(\ref{set}) and (\ref{vij}). Specifically, we calculate the following integrals
\begin{eqnarray}
\label{Inq3}
V_{1,1}&=&\int  {\rm dP}f_{eq}(p)\tau=\frac{n}{T},\nn\\
\quad V_{2,2}&=&\int  {\rm dP}f_{eq}(p)\frac{(u\cdot p-\frac{e}{n})^2(u\cdot p)}{T^3}=\frac{I^{(0)}_{30}-\frac{e^2}{n}}{T^3},\nn\\
V_{3,3}&=&V_{4,4}=V_{5,5}=\int  {\rm dP}f_{eq}(p)\frac{\tau(l\cdot p)^2}{T^2}=\frac{h}{T^2}.
\end{eqnarray}
For the massless case,
\begin{eqnarray}
\label{Inq4}
V_{1,1}=\exp(\xi)\frac{T^2}{\pi^2},\quad V_{2,2}=\exp(\xi)\frac{3T^2}{\pi^2},\quad
V_{3,3}=\exp(\xi)\frac{4T^2}{\pi^2},
\end{eqnarray}
where the thermodynamic integrals are calculated according to  Appendix.\ref{int}.

\clearpage

\end{appendix}
\bibliographystyle{apsrev}
\bibliography{ref}{}

\end{document}